# The Implementation of Arduino Microcontroller Boards in Science: A Bibliometric Analysis from 2008 to 2022

**Norbertus Krisnu Prabowo[1], Irwanto Irwanto[2]**
[1,2] Department of Chemistry Education,
Universitas Negeri Jakarta, Jakarta 13220, Indonesia
[1] norbertuskrisnu@gmail.com,
[2] Irwanto@unj.ac.id

**Abstract:** The name "Arduino" made its international debut in 2005, marking the age of Arduino as one of the most user-friendly and cost-effective microcontroller boards (MCBs) for novices. The science implementation of Arduino boards in automation, networking and data acquisition has been increasing steadily. This study provides a thorough Bibliometric analysis from 1122 papers focused on the Scopus database of published microcontroller research, from the first year the Arduino keyword appeared in 2008 until 2022. Various science articles indexed by Scopus and referring to the use of Arduino MCBs are selected. The Bibliometric analysis explores comprehensive and general key attributes that form a trend from the Scopus articles based on authors, titles, publication years, keywords, citations, affiliations, abstracts, funding information, and languages. The generated data is visualized and analyzed to find patterns that appear within the time span. This study found a significant increase in the number of articles on Arduino boards in Biology, Physics, Chemistry, Science, and STEM category of the paper. Despite using only the Scopus database, this study opens up to view the direction of the growing application of Arduino boards in Science. The use of Bibliometric analysis maps the scientific implementation of Arduino boards as an extensive guide for future collaborations in education and industry.

**Norbertus Krisnu Prabowo**
Department of Chemistry Education,
Universitas Negeri Jakarta, Jakarta 13220, Indonesia
norbertuskrisnu@gmail.com,

## 1. Introduction

The use of Arduino microcontroller boards (MCBs) has evolved steadily over the years, and it has been developed in many research areas. Various integrations in the research papers show a clear direction for the implementation of Arduino boards in science applications, ranging from education (Guven et al., 2022), health sensor technology (Bhardwaj et al., 2022), applied engineering (Loukatos et al., 2022) real-time monitoring (Janeera et al., 2021), Internet of Things (IoT) (Raju et al., 2020), homemade portable devices (Mariani et al., 2022) and many more. Many papers have reported the need for monitoring equipment for science experiments with Arduino boards (Christenson et al., 2022; Sudarmanto at al., 2022; de Vera et al., 2022). Unlike other microcontroller development platforms, an Arduino platform comes with a kit which can be used as an alternative to overcome many limitations of low-budget science research (Tran et al., 2022). This strengthens the disruptive era among researchers in terms of system automation and data acquisition in science applications. The implementation provides a new opportunity for innovation and complexity in the fourth industrial revolution domain (Bloem et al., 2014). The use of Arduino boards has shown an impact and contribution to the development of science.

The study in this paper is taken from the record of 421 journals, 613 conference proceeding articles, and other document types. There are 1122 publications from Scopus database regarding the science implementation of Arduino boards. Despite the multitude of publications available, the challenges associated with using Arduino boards and the skillset acquired by users when working with them have received relatively little attention. The use of sensor technology in microcontroller boards is claimed to give impacts in stimulating creative thinking, solving science problems, and innovative ideas allowing deep understanding of science under a low-cost hardware and software. The exploration of these parameters will give benefits for the research community to establish their future research when implementing Arduino boards in science. Therefore, a bibliometric study is required to unlock the potential research trend retrospectively. This motivated the initiation of the research. The earliest publications in this field came from two conference papers published (Buechley et al., 2008; Eliasz, 2009). The publications continue to grow. Hence, the time span for this study was taken from 2008 to 2022.

## 2. Rationale of the Study

There has been no previous bibliometric analysis of Arduino boards related with the implementation in Science and Education, which covers more than 10 years of Scopus database. In this study, a thorough bibliometric analysis was conducted in this domain. This study screened 1122 articles within the past 15 years. The updated patterns and trends are shown in this paper. This bibliometric study contributes a broad review of publications connected to Arduino boards, which were used in the field of Biology, Chemistry, Physics, Science, or STEM in 2008-2022. Therefore, this study is able to empower researchers to obtain a thorough overview, recognize a research gap, and synthesize a novel research idea in the area of Arduino boards and Science.

## 3. Objectives of the Study

This study explores the bibliometric variables mapping the emerging trends from the Scopus database. The objectives of the study focus to evaluate the publication and citation trends on Arduino-related science application throughout the year. From the top cited publications, we want to explore the challenges associated with using Arduino boards and the skillset acquired by users. We also want to find the most productive sources, universities, countries, the most influential authors, and the distribution of co-occurrence keywords contributed to the topic.

## 4. Literature Review

Microcontrollers are small electronic modules that are composed of one or more processors to execute specific tasks and control a system by obtaining data from the natural world stimuli (analogue) into the electronic world (digital) or vice versa (Sanchez & Canton, 2018). They work as a bridge between the two worlds. They are embedded systems that, when integrated into a device, they can manage and govern the functionalities and operations of the device. Microcontrollers are present in a wide range of devices including medical implants, industrial equipment, automotive engines, electronic appliances, and toys. Among the various systems available for the development board, Arduino has emerged as a hardware and software company that produces microcontroller-based Arduino boards since 2005. The use of Arduino boards, also known as Arduino Microcontroller Boards (MCBs), is increasingly growing to receive prominent attention from many researchers worldwide. The urgency for robotic innovation has been boosted in the current fourth industrial revolution (Elayyan, 2021; Lee et al., 2018).

The first Arduino board was introduced in 2005 to help novices to create electronic prototypes connecting the physical world to the digital world (DesPortes & DiSalvo, 2019). The word, Arduino, came from Italy, where the founders often gathered (Kondaveeti et al., 2021). The Arduino platform is an open-source system. The online community of users is available worldwide, providing support to the programming of Arduino in Python and C++ languages. Arduino boards have many different shapes and sizes, possessing diverse functionalities. Various common types of Arduino boards are presented in Table 1. The shape, physical dimension, processor, I/O capacity, and analogue input are examples of the important factors to be considered by users when prototyping using Arduino boards. A smaller dimension board is preferred as a component in RC cars or planes. With more digital input and output (I/O) pins, certain boards are more suitable as a machine controller in industry. The first Arduino board that utilizes a USB connector is Arduino Uno. This is perhaps the reason behind its popularity. The capacity and functionality of the boards can be extended using Arduino Shields, which are used by stacking on the top of the boards.

**Table 1: Various Common Arduino Boards**

| Boards | Dimensions (mm) | Microcontroller | Digital I/O | Analogue Input |
|---|---|---|---|---|
| Duemilanove | 68 x 53 | ATmega168 | 14 | 6 |
| LilyPad | diameter (50 mm) | ATmega168V | 14 | 6 |
| Nano | 18 × 45 | ATmega168 | 14 | 8 |
| Pro Mini | 34 x 18 | ATmega328P | 14 | 6 |
| Mega | 102 × 54 | ATmega2560 | 54 | 16 |
| Uno | 69 × 54 | ATmega1284 | 14 | 6 |
| Due | 102 × 54 | ATSAM3X8E | 54 | 12 |
| Leonardo | 69 × 54 | ATmega32U4 | 20 | 12 |
| Micro | 48 ×18 | ATmega32U4 | 20 | 12 |
| MKR1000 | 62 × 25 | ATSAMW25 | 8 | 7 |
| Yun | 68 x 53 | ATmega32u4 | 20 | 12 |

All of the boards can be equipped with electronics to communicate with sensors, LCD displays, and computers. Users can write codes into Arduino boards using Arduino IDE software (Pratomo & Perdana, 2017). The publication of the hardware and software is based on the Creative Commons License. This allows users to build, customize, improve, and distribute the assembled modules. They are many sensors and electronic parts that are compatible with Arduino. It can also be coupled with a chip radio transceiver, in-built Wi-Fi and Bluetooth, producing a potent Internet of Things (IoT) interconnectivity (Saini & Dutta, 2020). Merging these features in science applications can be beneficial.

## 5. Methodology

A. Data Collection

This study used bibliometric analysis to examine all Arduino articles related to science applications. The bibliometric analysis uses simple statistics to visualize and describe a large amount of data from scientific articles in a specific domain (Donthu et al., 2021a). It shows journal performances, collaboration patterns, research constituents, general evaluations, and interpretations in mapping the emerging trends in publications, languages, number of citations, authors, and types of articles, keywords, countries, and institutions (Donthu et al., 2021b).

A compilation of 1122 documents was obtained from the Scopus database. They have been enormously cited by 6729 other papers. It was equivalent to 5.99 citations per article. Scopus was used because it provides more extensive coverage of documents than other scientific databases in Natural Sciences and Engineering (Mongeon and Paul-Hus, 2016) Scopus is a world-class database that includes a high-quality publication database, comprising more than 7 thousand publishers. It comprises moe than 84 million records, 1.8 billion cited references, 17.6 million authors, 94.8 thousand affiliation profiles, and 33 thousand Science publications (Elsevier, 2022). Scopus unlocked a wide range of Science articles used in this paper. Therefore, it opened an easy access to the aggregated bibliographic data, which was relevant to the terminology of Arduino boards related to Science applications in Chemistry, Physics, Biology, or STEM.

**Table 2: Criteria For The Inclusion And Exclusion**

| Criteria |
|---|
| Inclusion Criteria (IC) |
| 1. The screening was limited to the title, abstract, and keywords of the documents (IC1) |
| 2. The documents must be written in the English language (IC2) |
| 3. The documents must focus on Arduino boards in Science Applications (IC3) |
| 4. All date of publications (IC4) |
| 5. All types of documents (IC5) |
| Exclusion Criteria (EC) |
| 1. The documents published in non-English language (EC1) |
| 2. The documents not related with Arduino boards in Science Application (EC2) |
| 3. Documents focused on Arduino boards in non-Science Application (EC3) |

The search produced 1115 documents after screening with the inclusion and exclusion criteria. Table 2 presents the details of the criteria. PRISMA, also known as, The Preferred Reporting Items for Systematic Reviews and Meta-Analyses, was used as a reference for the research protocol in this study (Moher et al., 2009). The flowchart is shown in Figure 1.

The metadata was subtracted on October 6, 2022 with no year limitation. All documents were considered including conference reviews, journal reviews, journals, books, book chapters, and editorials. It was intended to provide a thorough mapping of this topic. The articles was found in the Scopus database by using the following primary search: TITLE-ABS-KEY ("STEM") OR "Chemistry" OR "Physics" OR "Biology" OR "Science" AND "Arduino". When the metadata were checked, many articles were found related to Arduino boards in science applications. There were 10 countries and 11 languages. Articles written in two different languages were detected in the metadata (see Figure 1).

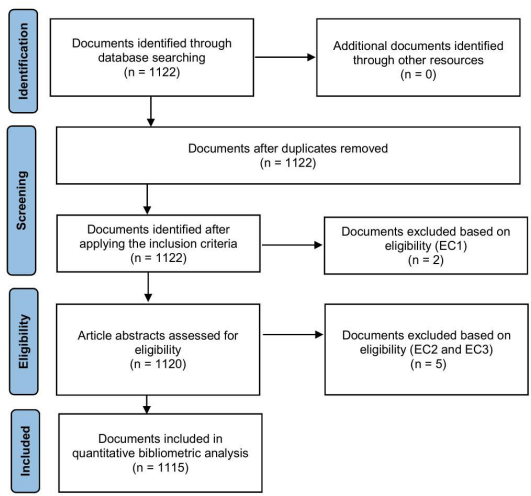

**Fig. 1: PRISMA Flowchart for the Bibliometric Analysis**

B. Data Analysis

The authors thoroughly screened the articles from the Scopus database by using the research protocol. The publications that meet the parameters were compiled. Data elicitation was conducted. The variables were classified to answer the research questions. From 1115 documents, 54.98% were proceeding papers (n=613), 37.76% were journal articles (n=421), and 7.26% were other document types (n=81). Both CSV (comma-separated values) and RIS (research information systems) files were downloaded from Scopus database. IBM SPSS Statistics version 26.0, Microsoft Excel, and Google Spreadsheet were used in the statistical analysis. An open-source software, VOS viewer Version 1.16.18, was used to visualize the bibliometric networks (Eck & Waltman, 2022). Data visualization was obtained, including the co-authorship, keyword co-occurrence, and citation analyses. VOS viewer was used due to user-friendliness, easy accessibility, and well-accepted bibliometric software (Eck & Waltman, 2022). Tables and visualization maps were constructed to be analyzed.

## 6. Results

The screened documents encapsulate that all relevant documents were published from 2008 to 2022. The trend in publications and citations every 5 years is presented in Table 3. The number of publications with the percentage was calculated from the 1115 documents that have passed the research protocol, while number of citations with the percentage was calculated from the total 6686 citations in the period of 2008-2022.

**Table 3: Five-year Comparison of Publications and Citations**

| 5-year period | The Number of Publications (The Percentage) | | The Number of Citations (The Percentage) | |
|---|---|---|---|---|
| 2008-2012 | 30 | (2.69%) | 864 | (12.92%) |
| 2013-2017 | 340 | (30.49%) | 3565 | (53.32%) |
| 2018-2022 | 745 | (66.82%) | 2257 | (33.76%) |

In Table 3, both the number of publications and citations increase significantly. From the first to the second 5-year period, the number of publications spikes more than ten-fold, while the number of citations quadruples. This indicates a considerable interest from the Science Community to utilize Arduino boards further. A significant increase is also observed from the second to the third 5-year period of publications and citations. However, it should be noted that the search period has yet to reach a full year in 2022, as the metadata was subtracted on October 6, 2022. This topic has attracted researchers' attention for the last 15 years, suggesting a potential topic for future research.

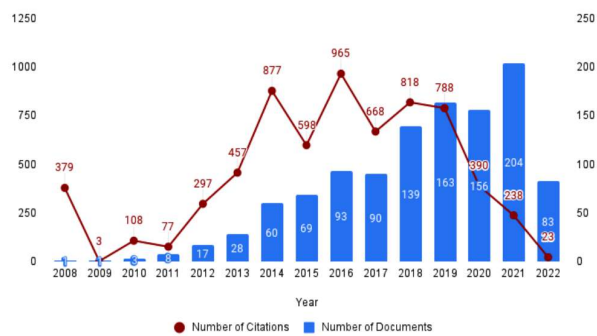

**Fig. 2: Distribution of Publications on Arduino board-related Science Application by Year**

Figure 2 represents the annual publications and progressive citations on Arduino boards related to science applications by year. The first publication on this topic appeared in 2008 with just one document. This document is a journal that utilizes the earlier type of Arduino, LilyPad Arduino, as a construction kit enabling novices to customize wearables and other textile artifacts. It was recorded by Scopus to have 379 citations, marking the entrance of Arduino boards in Science related applications. The growth of this topic's publications and citations in Figure 2 has fluctuated. The peak publication is observed in 2021, with 204 publications. The number of citations climaxes in 2016 with 965 citations.

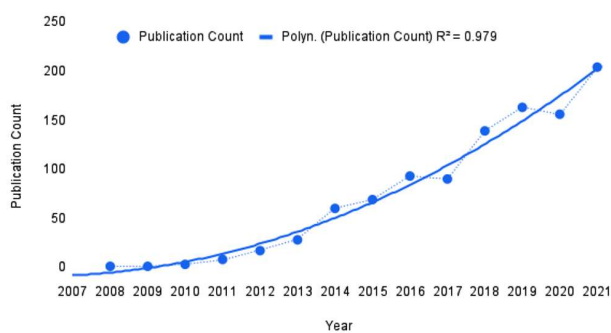

**Fig. 3: The Trend Analysis of Publication Counts**

Statistical computing was used to further uncover the emerging trend in the publications on the topic. A polynomial regression analysis was conducted using the independent variable (year) on the x-axis and publication count on the y-axis. The result is presented in Figure 3. Regression modeling is applied to fit the year records in 2008-2021. A polynomial regression curve with the $R^2$ value was integrated into Figure 3. The equation is calculated as follows, $y = 0.953x^2 - 3825x + 3840000$, $R^2 = 0.979$. The result shows a trend towards an exponential increase in the number of publications, strongly confirming the growth of enthusiasm for research on the topic. A positive coefficient of $x^2$ and a high value of $R^2$ are two indicators in the estimated regression model for more studies and research publications in the future (Donthu et al., 2021a).

Table 4 presents the top 10 most highly cited documents from 2008 to 2022. The article written by Buechley et al. in 2008 has the highest number of citations (C) and average citation per article (C/A). It was recorded in the Scopus database as the first journal that implemented Arduino board as a learning media in computer science education. While the article was published in April 2008, the type of Arduino incorporated in the research article was LilyPad Arduino, which was released a year before. In a short time, researchers were attracted to utilize Arduino boards. The second top-cited article, written by Kafai et al. in 2014, has 139 citations. Then, an article written by Ali et al. has 124 citations (C) and 17.71 (C/A) per year. The publications and citations continue to grow.

Table 4 also provides a concise review, which includes the type of Arduino boards used, research focus, level of participants, challenges and skills obtained. The intention behind the introduction of Arduino development boards is to expedite the process of application development. Therefore, Arduino currently has over 100 hardware products, including boards, shields, kits and other accessories. The results in Table 4 establish the key points when prototyping with Arduino boards. Educators and researchers might focus on the suitable characteristics of the Arduino boards when designing lessons, experiments, or systems.

From the total citations in this topic, the citation contribution from the top 10 articles in Table 4 is almost 20%. From 1115 publications, 1.79% from the screened documents have more than 50 citations, and 15.24% have more than 10 citations; 63.86% of publications earned at least one citation. Table 4 shows the trends in the top 10 most cited publications, which are related with the applications of Arduino boards in E-textile, Environmental Science, Neuroscience, and STEM Education (see Table 4). In this study, there were 464 different publishing venues detected from the screened 1115 documents. We observed that only 110 (23.71%) sources published 10 or more documents. In Table 4, the publishing venues include publications from Conference on Human Factors in Computing Systems (CHI), ACM Transactions on Computing Education (TOCE), Building and Environment Elsevier (BE), IEEE Frontiers in Education (FIE) Conference, Journal of Neuroscience Methods Elsevier (JNM), IEEE Computer Society (CS), ACM Special Interest Group on Computer Science Education (SIGCSE), Environmental Monitoring and Assessment Springer (EMA), Journal of Chemical Education American Chemical Society (JCE). In general, this is an indicator of the worldwide interest in implementing Arduino boards for Science.

**Table 4: Top 10 Most Cited Publications**

| Author(s) | Title, Arduino Board Used | Source | C | C/A | Research Focus, Participant Level / Age | Challenges when using the board | Skillset acquired by users |
|---|---|---|---|---|---|---|---|
| Buechley et al. (2008) | The LilyPad Arduino: Using Computational Textiles to Investigate Engagement, Aesthetics, and Diversity in Computer Science Education, Arduino LilyPad | CHI | 379 | 25.27 | Developed an expressive medium for textile-based construction kit that allows users, both school children and adults, to build and program their own wearable computers, aged 10-14. | Stitches were difficult to remove when incorporating the board. The round-shaped board challenged the participant's sew-ability. | Participant learned to invest in a period of careful engineering and design before embarking on constructing an e-textile. Simple programming & electronics skills were incorporated into the aspects of aesthetics and design, such as touch-sensitive wearables and handbags. |
| Kafai et al. (2014a) | A Crafts-Oriented Approach to Computing in High School: Introducing Computational Concepts, Practices, and Perspectives with Electronic Textiles, Arduino LilyPad | TOCE | 139 | 15.44 | Showcased the feasibility of bringing e-textiles into the high school CS classroom and its impact on students' perceptions of computing, aged 16–18. | The round-shaped board challenged participants to measure the length and direction of the wires thoroughly. Water proof materials needed to be considered to protect the electronics. | Participants learned to solve the short circuit problem, learned to align parallel circuits to avoid electronic topology problem, and learned to alter a faulty circuit design that incorrectly connected positive to negative. |
| Ali et al. (2016) | Open Source Building Science Sensors (OSBSS): A low-cost Arduino-based platform for long-term indoor environmental data collection, Bare-bones Arduino Uno (Pro Mini) | BE | 124 | 17.71 | Developed OSBSS platform & demonstrated the performance in an educational building, including air and surface temperatures, air relative humidity, human occupancy, light intensity, and $CO_2$ concentrations, University students. | The manual assembly of all the parts onto the board was time consuming, which involved multiple steps of soldering the circuit. | Users found there were errors in libraries or code downloaded from Arduino IDE (updating was required). Users built a system supporting future upgrades in hardware and software of the devices. |
| Sarik and Kymissis (2010) | Lab Kits Using the Arduino Prototyping Platform, Arduino Duemilanove | FIE | 103 | 7.92 | Described the design and Implementation of lab kits for a course that covers concepts related to science and modern display systems for both on-campus and remote students, University students. | Some students with limited prototyping or programming experience initially seemed "overwhelmed" by the platform. Introductory session was needed. The list of electronic parts, functions, and clear instructions must be provided. | Students learned to modify the existing code to utilize the color sensor and the RGB LED. Students could quickly diagnose problems with the board or the display. Students compared the brightness, contrast ratio, power consumption, and scalability of the different displays in a multi-segment EL display, a dot matrix inorganic LED display, and a graphic LCD. |
| Teikari et al. (2012) | An inexpensive Arduino-based LED stimulator system for vision research, Arduino Uno | JNM | 90 | 8.18 | Developed an inexpensive system to control light intensity for research applications in vision research and tested in rodent pupillometry, University students | For further embedded applications with Matlab, LabView, & etc, the Atmel-based performance may be insufficient | Users performed the use the Arduino Uno as a data acquisition device to read analog/digital inputs from external switches, buttons, joysticks, sensors, and providing signal out to control the LED drivers. |

 

**Table 4: Top 10 Most Cited Publications (continued)**

| Author(s) | Title, Arduino Board Used | Source | C | C/A | Research Focus, Participant Level / Age | Challenges when using the board | Skillset acquired by users |
|---|---|---|---|---|---|---|---|
| Peppler (2013) | STEAM-Powered Computing Education: Using E-Textiles to Integrate the Arts and STEM, Arduino LilyPad | CS | 88 | 8.80 | Integrated Arduino boards in Science, Technology, Engineering, Arts, and Mathematics (STEAM) in the digital fabrication techniques of e-textiles, aged 12-15. | Problems with design thinking process. Some designs failed to reflect original intentions. Some products have radical implications, which challenged existing cultural norms. | E-textiles were effective tools for broadening STEM disciplinary content. Students significantly increased their understanding of key circuitry concepts, circuit diagrams, current flow, circuit polarity, directionality, physical properties such as fabrics material and metallic-conductive thread. |
| Kafai et al. (2014b) | Ethnocomputing with Electronic Textiles: Culturally Responsive Open Design to Broaden Participation in Computing in American Indian Youth and Communities, Arduino LilyPad | SIGCSE | 83 | 9.22 | Proposed an approach to ethnocomputing that combines the teaching of computation and aspects of local culture, aged 12-15. | Student's difficulty was how to connect the cultural designs with the codes and the functionality aesthetically. | Students exhibited a high level of craft by executing straight and blanket stitches that were carefully sewn, as well as using complex patterns. Students' cultural identity was developed. |
| Saravanan et al. (2018) | Real-time water quality monitoring using Internet of Things in Supervisory Control and Data Acquisition (SCADA), Arduino Uno | EMA | 82 | 16.4 | Proposed a new SCADA system that integrates with the IoT technology for real-time water quality monitoring, University students | The version of Arduino Uno used did not have Wi-Fi connectivity. Additional GPRS module was involved in order to have a real-time monitoring. The board needed to be protected from water. | Users performed the integration of the sensor, processing, and communication module. The involvement of Arduino Uno was able to accelerate the speed of the SCADA system. The problem solving skill was acquired by designing an efficient water monitoring system via IoT. |
| Peppler and Glosson (2013) | Stitching Circuits: Learning About Circuitry Through E-textile Materials, Arduino LilyPad | JCE | 79 | 7.90 | Provided a foundation for integrating e-textile materials into standards-based practices in formal education systems and illustrated how this might be taught and assessed in the classroom, aged 7–12 | The conductive tip on LilyPad was on the front side only. Soldering was needed to ensure the contact between the conductive treads and the board. | The student's conceptual understanding of how electricity works was significantly increased. Instructors were able to identify the student's misconceptions about core circuitry concepts (current flow, connections and polarity). |
| He et al. (2016) | Integrating Internet of Things (IoT) into STEM Undergraduate Education: Case Study of a Modern Technology Infused Courseware for Embedded System Course, Arduino Yun & Uno | FIE | 78 | 11.14 | Described the challenges to design the IoT-based learning framework and proposed the effective learning approaches to overcome the challenges, University students. | Students needed more time to be familiar with the software. The STEM project in embedded system and IoT application were very challenging for novices. | Students gave positive feedbacks on the level of interest in IoT application development. Students retained better understanding and prototyping skills with hands-on projects, such as in platform security, code generation, networked applications, etc. |

**Table 5 : Top 10 Most Productive Sources**

| Source | A | C | C/A |
|---|---|---|---|
| Proceeding of The Institute of Electrical and Electronics Engineers (IEEE) | 100 | 201 | 2.01 |
| Journal of Physics: Conference Series | 87 | 149 | 1.71 |
| Physics Education | 52 | 175 | 3.37 |
| ACM International Conference Proceeding Series | 43 | 400 | 9.30 |
| ASEE Annual Conference and Exposition, Conference Proceedings | 33 | 67 | 2.03 |
| Journal of Chemical Education | 27 | 359 | 13.30 |
| Revista Brasileira de Ensino de Fisica* | 26 | 106 | 4.00 |
| IOP Conference Series | 24 | 65 | 2.71 |
| Advances in Intelligent Systems and Computing | 18 | 23 | 1.28 |
| Physics Teacher | 15 | 117 | 7.80 |

Notes: * Only publications in English language

Table 5 presents the top 10 most productive sources. The rank was ordered based on the number of total articles (A). Accordingly, these 10 sources have published 425 documents, accounting for 38.12% of the publication with several citations of 1660 (24.83%). The Institute of Electrical and Electronics Engineers Proceeding (IEEE) is listed as the top most abundant source on this topic, with 201 papers and 100 citations. The Proceeding IEEE has an impact factor of 14.91 (IEEE, 2022), an H-Index of 296, and a Q1 SJR value of 4.68 (Scimago Journal & Country Rank., 2022). However, ACM International Conference Proceeding Series holds the highest number of citations on this topic, with 400 citations. Association for Computing Machinery (ACM) was established in 1947 (ACM, 2022), while IEEE was established in 1963. The older journals presumably have well-established and produced a larger volume of peer-reviewed articles. Journal of Chemical Education holds the highest citation-to-articles ratio (C/A) presented in Table 5. This journal is a part of the well-known publications of the American Chemical Society. It has a 13.30 citation-to-articles ratio (C/A). This indicates that the articles published in this journal are impactful and capture the attention of many researchers. In general, the scope of the journals in Table 5 is extensive. This covers the area of education, engineering, computer, and electronics. The top 10 most productive sources are providing a diverse application of Arduino boards and continuing to give its influence globally.

Table 6 shows the most productive institutions. There are 10 top institutions in this topic. These institutions are located in Europe, Asia, Australia, South America, and North America. The list is based on the number of articles produced by the institutions.

There are 1066 institutions from all over the world detected in the Scopus database on this topic. The Universidad Nacional de Educación a Distancia (UNED) is the most productive institution in the top 10. The second position is Yeditepe University, with 9 papers and 26 citations. The third position is Universitas Negeri Jakarta, with 9 papers and 8 citations. University of Pennsylvania (US) has the highest citation-to-articles ratios (C/A) in the list. From the 2022 Scimago Institutions Ranking, the University of Pennsylvania is ranked number 32 from 4364 universities worldwide (Scimago Ranking, 2022). All of the institutions in Table 6 are credible Universities, with world ranks ranging from number 32 to 736 in the Scimago database. This illustrates the widespread influence and research interest of Arduino boards in many universities.

**Table 6: Top 10 Most Productive Institutions**

| Institution | Country | A | C | C/A |
|---|---|---|---|---|
| Universidad Nacional de Educación a Distancia (UNED) | Spain | 12 | 168 | 14.00 |
| Yeditepe University | Turkey | 9 | 26 | 2.89 |
| Universitas Negeri Jakarta | Indonesia | 9 | 8 | 0.89 |
| National Chiao Tung University | Taiwan | 8 | 139 | 17.38 |
| University of Pennsylvania | US | 8 | 319 | 39.88 |
| Utah State University | US | 8 | 98 | 12.25 |
| North South University | Bangladesh | 8 | 73 | 9.13 |
| National Taiwan Normal University | Taiwan | 7 | 38 | 5.43 |
| University of Szeged | Hungary | 7 | 29 | 4.14 |
| Universidade Federal de Mato Grosso Do sul | Brazil | 7 | 27 | 3.86 |

Table 7 depicts the most productive countries in this topic. There are 61 different countries contribute to this topic, including European countries, Asian countries, North American countries, African countries, and Oceania countries. More than 60% of the publications come from the top 10 most productive countries. In Table 7, the US ranks first with 199 publications, followed by India (113), Brazil (84), and Indonesia (72). The number of articles published by the top two countries gives the most significant gap. This reflects the prominent position of the country on this topic. The US dominates the number of citations and has the highest citation-to-articles ratio (C/A). The country with the second highest citation-to-articles ratio (C/A) is Spain, followed by Taiwan (7.20), India (5.69), and Germany (4.34).

**Table 7: Top 10 Most Productive Countries**

| Country   | A   | %     | C    | C/A   |
|-----------|-----|-------|------|-------|
| US        | 199 | 17.85 | 2201 | 11.06 |
| India     | 113 | 10.13 | 643  | 5.69  |
| Brazil    | 84  | 7.54  | 339  | 4.04  |
| Indonesia | 72  | 6.46  | 165  | 2.29  |
| China     | 43  | 3.86  | 111  | 2.58  |
| Spain     | 42  | 3.77  | 354  | 8.43  |
| Italy     | 42  | 3.77  | 118  | 2.81  |
| Taiwan    | 35  | 3.14  | 252  | 7.20  |
| Germany   | 29  | 2.60  | 126  | 4.34  |
| Greece    | 28  | 2.51  | 89   | 3.18  |

Table 8 presents the top 10 most prominent authors on the topic based on the number of articles published. The most productive author is M. Castro, with 11 publications, followed by others, as listed in Table 8. The citation-to-article ratios (C/A) are relatively higher for the papers written by authors in European countries. The paper written by Kafai has the highest citation-to-article ratio (C/A), with 22.5. This value is two times higher than Castro's citation-to-article ratios (C/A). This is an indication of an influential author in the field of Arduino MCBs and Science. Most of the authors in the list have high author-level metrics (H-index). Researchers with high H-index are easier to predict in terms of their number of publications and average citations per paper (Ayaz et al., 2018). In general, all prominent authors in Table 8 strongly indicate the research specialization in Electrical and Computer Engineering, affiliated with the top universities across the globe (Spain, United States, Taiwan, Brazil, Hungary, Morocco, and Turkey).

**Table 8 : Top 10 Most Influential Authors**

| Author          | Affiliation                                                 | Country | A  | C   | C/A   |
|-----------------|-------------------------------------------------------------|---------|----|-----|-------|
| Castro, M.      | Universidad Nacional de Educación a Distancia (UNED)        | Spain   | 11 | 130 | 11.82 |
| Kafai, Y.B.     | University of Pennsylvania                                  | US      | 8  | 180 | 22.50 |
| Plaza, P.       | Universidad Nacional de Educación a Distancia (UNED)        | Spain   | 8  | 103 | 12.88 |
| Sancristobal, E.| Universidad Nacional de Educación a Distancia (UNED)        | Spain   | 8  | 103 | 12.88 |
| Urban, P.L.     | National Chiao Tung University                              | Taiwan  | 8  | 118 | 14.75 |
| Carro, G.       | Universidad Nacional de Educación a Distancia (UNED)        | Spain   | 7  | 78  | 11.14 |
| Goncalves, A.M.B.| Universidade Federal de Mato Grosso do Sul                 | Brazil  | 8  | 23  | 2.88  |
| Mingesz, R.     | University of Szeged                                        | Hungary | 7  | 18  | 2.57  |
| Ouariach, A.    | Mohammed I University                                       | Morocco | 7  | 16  | 2.29  |
| Çoban, A.       | Yeditepe university                                         | Turkey  | 7  | 5   | 0.71  |

Figure 4 shows the collaboration between authors and co-authors. The minimum number of documents and citations of an author was first set to 1 in the VOSviewer. The co-authorship mapping analysis must show a collaboration of two authors or more in a joint research publication. Therefore, there are 3370 authors detected in the Scopus database on the topic, and only 2320 authors met the thresholds. The total strength of co-authorship provides links. These links with other authors was calculated. The authors with the greatest total link strength were selected. Most of the authors are not connected to each other. However, VOSviewer identified the most extensive set of connected authors consisting of 26 authors. These 26 authors are well connected to form two groups with 177 links. Figure 4 describes the visualization network of the 26 collaborated authors. A node in Figure 4 represents the individual author. The size of the nodes demonstrates the number of co-authored publications by the author—the more publications, the larger the size of the nodes (Eck & Waltman, 2022). Two clusters are visualized in different colors. The authors from the two clusters are working collaboratively. The red cluster shows 17 authors, while the green cluster comprises 9 authors. The finding in Figure 4 shows an active collaboration network among authors. We observed that Ruyle has collaborated in joint publications with many authors. As an active author, Ruyle provides the highest total link strength (TLS=19). The two most prominent nodes are J. E. Ruyle and C. E. Davis connecting the two clusters together. They have higher numbers of co-authored publications than their peers in Figure 4. Both authors are affiliated with the University of Oklahoma, US. They frequently collaborated with other researchers from the University of California, Ohio University, Cornell University, Calvin College, and University of Nevada. Each of the institutions provides 11 TLS. All of the institutions are located in the US.

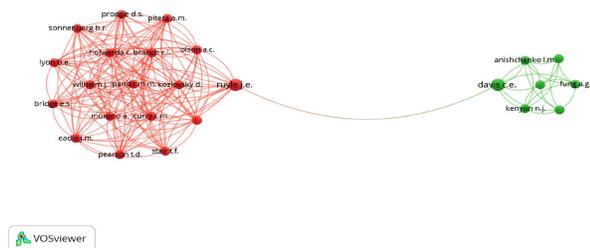

**Fig. 4: The Visualization Linkage in the Author Co-authorship**

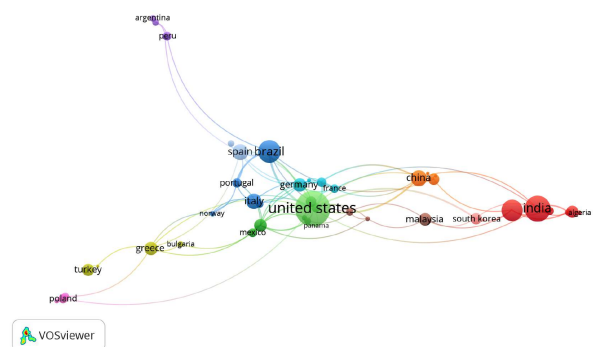

**Fig. 5: The Visualization Linkage of the Country Co-authorship**

The linkage among country co-authorship is presented in Figure 5. The parameters are set to have the minimum number of documents of a country as 1 and the minimum number of citations of a country as 1. Consequently, there are 51 of 61 countries met the thresholds. VOSviewer detected 12 clusters with TLS=128. The most prominent nodes are the US, India, Brazil, and China, indicating the significant contribution of the countries to the co-authored publications. As presented in Figure 5, the US, Spain, India, Brazil, and China have the most prominent nodes, reflecting the high productivity of the countries. In summary, these countries are the significant contributors to the publications involving Arduino boards in science applications.

The co-occurrence analysis was conducted using VOSviewer to produce a visualization map, as presented in Figure 6. It describes all of the keywords in the titles, abstracts, author keywords and index keywords. The related keywords were determined based on the number of publications in which they occur together (Eck & Waltman, 2022). Co-occurrence analysis demonstrates both author and index keyword occurrence in the articles. The unit analysis was selected for all keywords with a whole counting method. The minimum number of occurrences of a keyword was set to 5, and 394 keywords from the total 7738 keywords met the threshold. VOSviewer calculated and determined the total strength of co-occurrence links with other keywords.

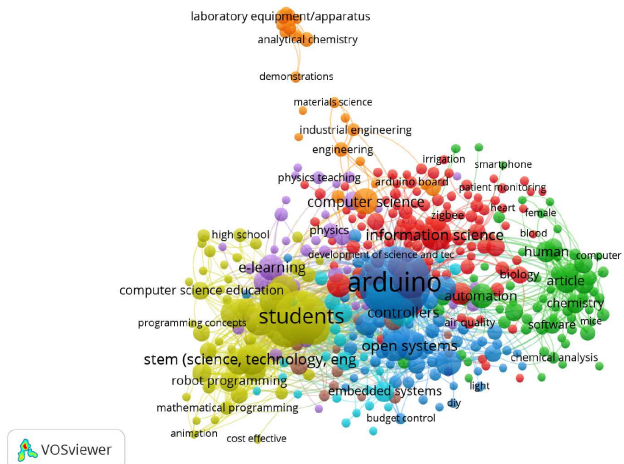

**Fig. 6: The Keyword Network Visualization**

The total link strength (TLS) was counted as 21994 with 11431 links in Figure 6. There are 8 clusters identified from 394 total keywords. The blue cluster having the most frequent keyword, Arduino, is located in the center of the map in Figure 6. There are 59 other keywords linked to it. The blue cluster is tight to several keywords, such as 3D-printing, cost-effectiveness, and LabVIEW. The blue cluster's total link strength (TLS) is 1955 with 374 occurrences. The brown cluster connects keywords, such as engineering education, computational thinking, problem solving, robotics, science education, and e-textiles. The brown cluster contains 275 links, 200 occurrences, and a TLS of 1561.

Table 9 displays the top 10 most frequent keywords. The author keywords hold 33.33% of the total keywords detected by VOSviewer. We observed the smaller keyword clusters, such as budget control, laboratory equipment, industrial engineering, smartphones, thermodynamics, green chemistry, genetics, and information science. We predict that the smaller keyword clusters are able to grow more significantly in the future. They open many gaps in research opportunities and provide new directions for future applications. The keywords (Arduino and Science) have established a strong relationship.

**Table 9: Most Frequently Used Keywords**

| Rank | Keywords | Occurrence | TLS |
|---|---|---|---|
| 1 | Arduino | 374 | 1955 |
| 2 | Engineering Education | 107 | 933 |
| 3 | Education | 83 | 743 |
| 4 | Robotics | 79 | 674 |
| 5 | IoT* | 93 | 672 |
| 6 | Microcontrollers | 78 | 611 |
| 7 | STEM** | 65 | 596 |
| 8 | Open-source software | 56 | 541 |
| 9 | Controllers | 55 | 470 |
| 10 | Information Science | 69 | 433 |

Notes: * Internet of Things, **Science, Technology, Engineering, and Mathematics

## 7. Discussion And Conclusion

This study assessed the research trends associated with Arduino boards, specifying the implementation of Chemistry, Biology, Physics, or Science using bibliometric analysis. The data were retrieved from 2008 to 2022. There were 1122 documents collected from the Scopus database. The data were reduced to 1115 documents using the inclusion and exclusion criteria in Table 2. The VOSviewer software was utilized for further analysis of the 1115 documents. The enterprise, Arduino, has been manufacturing microcontroller-based boards since 2005. The first worldwide products (Arduino Diecimila & Arduino LilyPad) were released in 2007. However, the first paper related to the topic appeared in 2008. It takes about one year for the first paper in the Scopus database to emerge. The scientific community needed time to embrace how to use Arduino boards with the codes, how to use it to solve science problems, and how to relate the science concepts in the methodology. From the research trends and keyword analysis, researchers have understood the characteristics of Arduino boards connected to science applications over time. The characteristics are user-friendly, cost-effective, and open-source (Cressey, 2017). It has captured the attention of many researchers.

Starting in 2009, the papers related to this topic increased gradually, with the highest publication peak in 2021, as presented in Figure 2. In 2021, there were 204 publications and 234 citations. These values are proportional to 18.30% of the total publications and 3.5 % of the total citations. In this current study, the annual publication trend shows an exponential growth in the number of documents published annually, which correlates to Price's law in the theory of bibliometric distributions (Price, 1963). This describes the tendency of an increasing number of publications on a specific topic. The polynomial regression curve also suggests an increasing trend with the R2 value of 0.979 in Figure 3. This finding is consistent with the previous bibliometric studies (López-Belmonte et al., 2020; Soares et al., 2019; Trento et al., 2020). Specifically, the number of publications and citations dramatically increased from 2013 to 2016. The highest number of citations in this study was 965 citations in 2016. The number of citations was gradually descending after 2016. The analysis using the 5-year window in the number of citations in Table 3 supports this finding. This was counterintuitive, as the number of citations indicates the quality of a publication (Patterson & Harris, 2009). The "quick impact" journals are cited due to the influence of the previous outstanding results published 2 years before (Campanario, 2011). However, the phenomenon of a descending trend in the number of citations is accepted as the older publications have more chance to accumulate citations (Aksnes et al., 2019).

In Table 4, the several top authors are influenced by the works of Buechley and stated them as references. It suggests that the top-rated publications have higher concentrations among their contributors, and frequently cited publications are more likely to be cited again. According to Lotka's law of scientific productivity, this is very often observed among high quality publications. Lotka called this phenomenon "success breeds success" (Daniel, 1997). The top 10 most cited documents are influential publications in implementing Arduino boards for Science. From the top publications related with education and e-textiles, both of male and female learners manifested great enthusiasm and motivation in prototyping using Arduino boards (Peppler, 2013). It engaged the participants to further explore the topics in computer science, electrical engineering, and STEM (Kondaveeti et. al., 2021). The use of Arduino board kit was effective when learners were allowed to make a large number of mistakes and instructors should do less scaffolding or intervening (He et al., 2016; Peppler & Glosson, 2013). The collaboration in designing, programming, and prototyping plays an important role in the stages of using Arduino software and hardware. Learners retained better skills using a hands-on and team-oriented learning approach (He et al., 2016). In most of the cases in Table 4, some novices commented that the incorporating a microcontroller board in a problem-solving project was very challenging. The takeaway from this finding is that instructors may begin with a diagnostic test in the class to screen the student's initial understanding in programming and electronics. An introductory session to familiarize the name of the electronic parts and functions, such as jumper cables, breadboards, pins, etc, will be helpful for novices. This may include the introduction to code writing in Arduino IDE.

The shapes of Arduino boards provide an effect in the functionalities. It also gives certain obstacles to the users when incorporating them in a project. For example, the round-shaped LilyPad gave more flexibility in the directions of the wirings for e-textile projects (Peppler & Glosson, 2013). However, users needed to carefully measure the length of the connecting wires and the soldering strength (Kafai et al., 2014a; 2014b), as the pins were relatively far from each other. Users have been reported using alligator clips for a temporary solution (Buechley et al., 2008). Arduino boards also show a strong connectivity as a data acquisition tool. This works with several interfaces, such as PLX-DAQ Excel, LabVIEW, and MATLAB (Nichols, 2017). The process of importing data directly from the Arduino boards is real-time, allowing further data presentation, plotting, and analysis. However, for complex setups with more embedded applications, the Atmel microcontroller's performance may be insufficient. Teikari et al. (2012) advised that the use of Arduino Uno board in the complex neuroscience research could be complemented with other compatible boards, such as BeagleBoard (http://beagleboard.org/) and Raspberry Pi (http://www.raspberrypi.org/). These boards have more computing power and compatibility with Linux system due to the ARM microcontroller, but they have less input/output pins compared to the Arduino Uno board.

Learning intervention using Arduino projects was reported to give a significant increase to the students' STEM academic achievement and perceptions towards STEM from a meta-analysis study conducted in K-12 and post-secondary classrooms (Fidai et. al., 2020). A mixed method research by Peppler and Glosson in 2013 has indicated significant gains in working circuits ($t(16) = 4.76, p < .001$ (two-tailed)), current flow ($t(16) = 3.34, p < .005$ (two-tailed)), connections ($t(16) = 2.31, p < .04$ (two-tailed)), and polarity ($t(16) = 4.74, p < .001$ (two-tailed)). The research was conducted using pre/post-test assessments at 0.05 level of confidence with ages ranging from 7 to 12 years. Current multiple studies that implemented Arduino projects have shown a positive improvement in the user's problem solving skill (Felicia et al., 2017), creative thinking skill (Hsiao et al., 2019), and the academic areas of biology, mathematics, and science (Saez-López et al., 2019).

The involvement of Arduino hardware and software in STEM education appear to be significant over the years, as presented in a high TLS in Table 9. The implementation of Arduino boards provides a way for educators to promote IoT, Artificial Intelligence, and 3D modeling. In the last decade, many website communities, such as https://www.instructables.com/ and https://www.tinkercad.com largely used and cited in the top publications (Walkowiak & Nehring, 2016; Darji et. al., 2022) are as they provide helpful guides for coding, designing, and prototyping with Arduino boards. Strengthening these findings, the leading publications exhibit a strong connection with remote learning and experiments. Two articles utilized Arduino boards in hands-on activities for remote students and take-home lab exercises (Sarik & Kymissis, 2010; Peppler, 2013). These teaching strategies are highly beneficial during the Covid-19 pandemic (Dewilde & Li, 2021; Chichekian et al., 2022; Liu et al., 2021; Papadimitropoulos et al., 2021; West et al., 2021), as many educators are challenged to adapt to remote learning. In general, the leading sources cover the area of Engineering, Education, Physics, Chemistry, Biology, Education, and Computers. There are growing opportunities for research on this topic. The keyword analysis by VOSviewer has revealed the trend and the direction of the growth of this topic. They lead to the fulfillment of the fourth industrial revolution, the extensive use of the Internet of Things (IoT), and the demand for the 21st-century learning skills in education. However, the wide application of Arduino boards requires time for technological readiness (Dachyar et al., 2019). The current studies show Arduino boards in Agriculture (Revathy et al., 2022), Hospital, and Health Industries (Pantawane et al., 2022), marking the continuation of research in implementing Arduino boards for Science.

Most of the highly productive sources are European institutions. However, the top producing countries are more spread out, as also reported by Soares et al. (2019). In Figure 5, the visualization network of the country co-authorship shows common global interests shared within this topic. In contrast to that finding, only 26 authors met the thresholds and were well connected (see Figure 4). The total number of authors detected by VOSViewer was 2320 authors. Therefore, only 1.12% of the authors have collaborated between institutions. It displayed that most of the researchers were still working separately. Future researchers are highly encouraged to collaborate more frequently among institutions in this field.

Concerning the citation-to-article ratio (C/A) of the top sources, the Journal of Chemical Education was found to have the highest ratio, with a C/A of 13.30. Fages (2010) stated that the articles with high citation-to-article ratios are more frequently read and cited than others. Supporting this finding, Fages has provided quantitative proof using OLS regressions on readability scores from higher-ranked journals. This is one of the parameters of impactful journals. The Journal of Chemical Education is a prestigious peer-reviewed academic journal published by the Chemical Education Division of the American Chemical Society. It is one of the world's largest scientific societies, with over 151,000 memberships and represents more than 140 countries. It attracts many researchers. Its publications division produces over 60 scholarly journals (ACS, 2022).

The most productive countries are the countries with strong TLS links, showing frequent collaboration among authors in the network of the country co-authorship. The US contributed 199 publications and 2201 citations in this topic. Most of the implementation of Arduino boards in the US publications are related to Education. We predict that countries that invest more in microcontroller board technology will benefit in a number of ways, particularly in Education. According to Ivanović and Ho on bibliometric analysis, the US is the absolute leader in this area (Ivanović & Ho, 2019). India followed it with 113 publications. Among these publications, VOSviewer detected 108 Indian authors who worked collaboratively. Spain is also listed as one of the top 10 most productive countries, institutions, and prominent authors in this topic. M. Castro, P. Plaza, E. Sancristobal, and G. Carro are prominent authors from Spain with relatively high citation-to-article ratios. All of them are in the same institution, Universidad Nacional de Educación a Distancia (UNED). In Figure 5, Spain, Chile, and Argentina are in the same network of the country co-authorship. This signifies the major international collaboration among these three countries. They are Spanish-speaking countries and share solid cultural bonds providing flexibility in communication and research (Ardila, 2020).

Regarding the influential articles, reputable publishers (e.g. Elsevier and Springer) contribute to the high number of citations. Articles written by Ali et al. (2016), Teikari et al. (2012), Saravanan et al. (2018), Peppler and Glosson (2013) are published in those reputable publishers. Those are the top-cited articles in this field. Springer and Elsevier have high credibility in scientific publications and have a robust research community (Mongeon & Paul-Hus, 2016). There are five professors in the list of the top 10 prominent authors. First, M. Castro, Professor of Electronic Technology Engineering at UNED, is an expert in Computer Science, ICT, and Remote Laboratories. Second, Yasmin B. Kafai, Professor of Learning Sciences at the University of Pennsylvania, is an expert in Computing Education and Electronic Textiles. Third, Pawel Urban, Associate Professor at National Tshing Hua University, focuses his research on Analytical Chemistry and Chemical Biology. Fourth, A.M.B. Goncalves, Associate Professor at the Universidade Federal do Mato Grosso do Sul, is a material science and physics expert. Fifth, Robert Mingesz, Associate Professor at the University of Szeged, is an expert in quantum cryptography, the Internet of Things (IoT), and STEM.

In the grand scope, Arduino boards have impregnated the science pedagogy in STEM and laboratory instruction. From the 1115 articles used in the bibliometric analysis, 67.1% strongly relates with STEM and Education. The rest of the articles focusing on Physics, Biology, and Chemistry respectively are 12.1%, 6.3%, and 7.9%. It has been supporting the rise of the new term ChemDuino (a portmanteau of Chemistry and Arduino) since 2015 (Kubínová & Šlégr, 2015; Walkowiak & Nehring, 2016; Küçükağa et al., 2022). In this study, we found that the general practice of applying Arduino hardware and software has become well-accepted in many fields of Science, including Chemistry Biology and Physics. Thus, the network of keywords between Arduino boards and Science are interrelated via Engineering, Robotics, and STEM.

Arduino boards continue to challenge researchers and educators for creativity and innovation to solve common science problems, such as in sensor automation, networking and data acquisition. Surprisingly, the percentage of articles connected to the embedded health measurement systems using Arduino boards is about 6.6% from the total of 1115 articles screened. We predict that the percentage of publications in healthcare services will continue to grow in the future, as indicated by the impactful breakthrough in utilizing Arduino boards for remote health monitoring via Internet of Things (IoT) (Odusami et. al., 2022) and biomedical instrumentations (Salama et. al., 2022).